# NiO$_x$ passivation in perovskite solar cells: from surface reactivity to device performance

*John Mohanraj,[1] Bipasa Samanta,[2] Osbel Almora,[3,4] Renán Escalante,[4] Lluis F. Marsal,[3] Sandra Jenatsch,[5] Arno Gadola,[5] Beat Ruhstaller,[5] Juan A. Anta,[4] Maytal Caspary Toroker,[2,6] Selina Olthof[1]\**

[1]Department of Chemistry, University of Cologne, Greinstrasse 4–6, Cologne 50939, Germany

[2]Department of Materials Science and Engineering, Technion–Israel Institute of Technology, Haifa, 3600003 Israel

[3]Departament d'Enginyeria Electrònica Elèctrica i Automàtica, Universitat Rovira i Virgili, 43007 Tarragona, Spain

[4]Center for Nanoscience and Sustainable Technologies (CNATS). Department of Physical, Chemical, and Natural Systems, Universidad Pablo de Olavide, Sevilla 41013, Spain

*[5]*Fluxim AG, Katharina-Sulzer-Platz 2, 8400 Winterthur, Switzerland

[6]The Nancy and Stephen Grand Technion Energy Program, Haifa 3200003, Israel

**TOC** 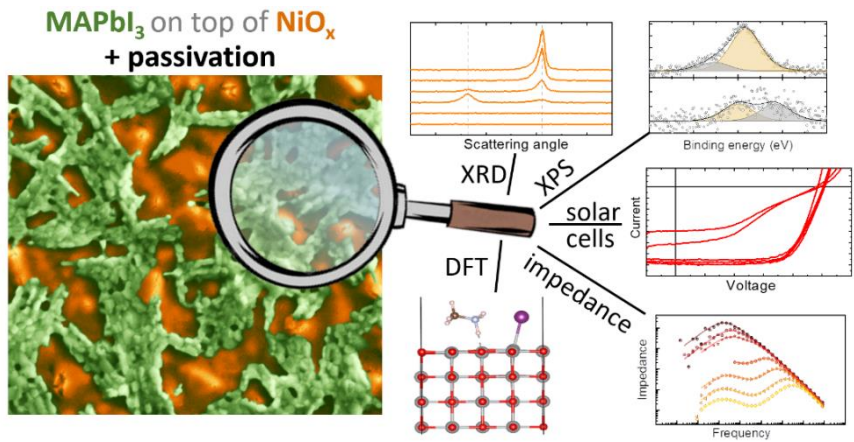

## Abstract

Non-stoichiometric nickel oxide (NiO$_x$) is the only metal oxide successfully used as hole transport material in p-i-n type perovskite solar cells (PSCs). Its favorable opto-electronic properties and facile large-scale preparation methods are potentially relevant for future commercialization of PSCs, though currently low operational stability of PSCs containing NiO$_x$ hole transport layers are reported. Poorly understood degradation reactions at the interface to the perovskite are seen as cause for the inferior stability and a variety of interface passivation approaches have been shown to be effective in improving the overall solar cell performance. To gain a better understanding of the processes happening at this interface, we systematically passivated possible specific defects on NiO$_x$ with three different categories of organic/inorganic compounds. The effects on the NiO$_x$ and the perovskite (MAPbI$_3$) were investigated using x-ray photoelectron spectroscopy (XPS), X-ray diffraction (XRD), and scanning electron microscopy (SEM) where we find that the structural stability and film formation can be significantly affected. In combination with Density Functional Theory (DFT) calculations, a likely origin of NiO$_x$-perovskite degradation interactions is proposed. The surface passivated NiO$_x$ was incorporated into MAPbI$_3$ based PSCs and its influence on overall performance, particularly operational stability, was investigated by current-voltage (*J-V*), impedance spectroscopy (IS), and open circuit voltage decay (OCVD) measurements. Interestingly, we find that a superior structural stability due to an interface passivation must not relate to high operational stability. The discrepancy comes from the formation of excess ions at the interface which negatively impacts all solar cell parameters.

# 1. Introduction

Perovskite solar cells (PSCs) have crossed the 25% power conversion efficiency (PCE) benchmark in the lab scale, fulfilling one of the prerequisites for commercialization.[1] As important next steps, researchers focus on the scalability of the devices into modules without compromising the efficiency and strive to improve the operational stability. With these goals in mind, PSCs with an inverted p-i-n device architecture are gaining interest due to their facile fabrication methods and compatibility with existing industrial processing procedures.[2] On the n-side of the device stack, $C_{60}$ fullerene and its derivatives (such as PCBM) are proven to be efficient as electron transport materials, though some voltage loss remains.[3] On the p-side, poly[bis(4-phenyl)(2,4,6-trimethylphenyl)amine] (PTAA) shows remarkable performance comparable to other hole transport materials employed in n-i-p device stacks.[4] However, despite its effectiveness, upscaling this polymer in large quantities would be cost and energy expensive, which hinders large scale processing. Further, the thermal stability of the PTAA/perovskite interface has also been questioned, which could limit the overall operational stability of the PSCs.[5]

Besides PTAA, thin layers of non-stoichiometric nickel oxide ($NiO_x$), prepared by a variety of methods such as solution processing, sputtering, and atomic layer deposition techniques, have also been explored as hole transport layer (HTL) in PSCs.[6] $NiO_x$ is a promising hole extraction layer due to its inherent p-type nature, wide bandgap, high hole mobility, and high chemical and thermal stability. At the same time, the upscaling of $NiO_x$ is unproblematic and it can even be processed and used on flexible substrates. These advantageous properties led to a wide-spread use in other fields of emerging optoelectronics, such as dye sensitized / organic solar cells and light emitting diodes.[7–12]

$NiO_x$ therefore appears to be a feasible alternative to be employed in commercial perovskite-based devices. However, the use of bare $NiO_x$ in contact with the perovskite layer has met unexpected challenges with efficiencies typically below 20%,[6] while the previously mentioned PTAA-based devices show higher efficiencies (24%).[4] In general, the interfaces between metal oxides and perovskites are known to be problematic, due to redox reactions that can take place, leading to a decomposition of the perovskite film.[13–18] $NiO_x$ is also known for its rich surface chemistry and high catalytic activities and can exhibit a variety of surface components such as $Ni(OH)_2$, $NiO(OH)$, and $Ni_2O_3$.[10] Therefore, this surface could present a bottleneck for commercialization, as the various surface species are claimed to cause the degradation of the perovskite absorber layer, resulting in interfacial charge carrier recombination and a loss in PSC performance over time.[19]

To circumvent the degradation at the interface and overcome the PSC performance issues related to $NiO_x$ surface defects, a wide range of surface modification techniques have been implemented in the past. For example, a simple UV-Ozone treatment of the $NiO_x$ layer is reported to affect the surface defects, enhance the surface wettability and its electronic properties, and with it the performance and stability of the PSCs.[20] Similarly, conductive/insulating polymeric thin films deposited at the interface spatially separate the $NiO_x$ from the perovskite and thereby help to alleviate the degradation issues.[21,22] Extrinsic doping of $NiO_x$ with metal ions such as Li,[23] Cs,[24] and Cu [25,26] has also been shown to improve the performance of the $NiO_x$-based PSCs. Another widely employed method is targeted selective surface passivation, in which specific organic molecules [27–31] as well as alkali halides [32,33] have been deposited on $NiO_x$ to passivate surface defects.

The various studies claim that surface defects such as Ni vacant sites, surface hydroxyl groups, or high valent surface Ni species (Ni > 2+) have been passivated, which otherwise would have acted as catalytic degradative reaction centers, leading to charge carrier recombination. Interestingly, the employed organic molecules fall under a wide spectrum including acids[27–29] as well as bases,[30,31] and all of them appear to improve the device performance. Recently, Boyd et al. suggested that the use of a slight excess of organic ammonium halide precursor can passivate the reactive Ni species and improve the open-circuit voltage ($V_{oc}$) as well as the operational stability of devices with NiO$_x$ HTLs.[34] They claimed that high valent Ni species can oxidize the halides and abstract a proton from the precursor ammonium halides, leading to degradation of the perovskite at this interface.

The varying chemical nature of the mentioned passivating materials makes it difficult to develop a coherent understanding of the nature of the reaction centers in NiO$_x$, as well as the underlying mechanisms of degradation/passivation. For example, claims regarding the effectiveness of both UV-ozone treatment and surface passivating techniques are conflicting. Upon treating NiO$_x$ layers with UV-ozone, the density of high valent Ni components is found to increase, which was suggested to be beneficial for devices.[20] On the contrary, some of the surface passivation techniques are claiming to target and remove such high valent Ni components, which is also reported to achieve high efficiency and stability.

All these successful observations on surface passivation of NiO$_x$ leave many open questions regarding the underlying mechanism. It is unclear whether the observed improvements in device performance and stability are indeed a result of NiO$_x$ surface passivation or due to any other side effect, for instance related to a manipulation of electronic/mobile ion concentrations at the NiO$_x$-perovskite interface. Moreover, the insertion of an interlayer removes the direct contact between NiO$_x$ and the perovskite film. Furthermore, convincing evidence is missing on the chemical interaction between the perovskite and the organic passivating materials, which are mostly Lewis bases in nature. This puts into question the passivating process itself and it is uncertain whether these unbound or loosely bound passivating materials remain on the NiO$_x$ during the fabrication of the device. These open questions highlight the need for further studies, focusing on understanding: (i) the origin of degradative interactions in NiO$_x$, (ii) the nature of interactions between the surface passivating materials and NiO$_x$ surface components, as well as (iii) their influence on the perovskite light absorber, device performance, and stability. The research presented here is meant to provide a step in that direction.

In this work, the interfacial interactions between solution processed NiO$_x$ and the perovskite MAPbI$_3$ are revisited by using X-ray photoelectron (XPS) and X-ray diffraction (XRD) techniques together with scanning electron microscopy (SEM) to investigate the perovskite stability issues. Next, we choose three different classes of surfaces passivating materials, targeting specific reaction sites on NiO$_x$, to perform a systematic investigation regarding the surface interactions. The relative changes in surface chemical composition of NiO$_x$, as well as the resulting impact on the stability of MAPbI$_3$ perovskite films, are tested. The results are correlated with density functional theory (DFT) calculations, in which NiO$_x$ surface passivation experiments are mimicked and the values of the reaction free energy for perovskite precursors degradation are compared. Based on these results, we propose a plausible reaction center for degradative interactions and a corresponding mechanism for perovskite degradation. Finally, we incorporate surface treated NiO$_x$ HTLs in MAPbI$_3$-based PSCs and investigate their influence on solar cell characteristics and device operational stability. Further, the impact of surface modified NiO$_x$ on photogenerated charge carrier recombination events and perovskite ionic properties in PSCs are studied by open-circuit voltage decay (OCVD) and impedance spectroscopy (IS) measurements. These complementary results reveal the significant role of NiO$_x$ surface modifications, which not only affect the film formation of the perovskite but also its electronic and mobile ion charge carrier concentration. Overall, we find that the choice of surface treatment can play a crucial role in determining the performance as well as the operational stability of PSC devices.

## 2. Results and discussion
### 2.1. Material characterization
#### 2.1.1. MAPbI$_3$ stability on bare NiO$_x$

In order to investigate the interactions between NiO$_x$ and the perovskite MAPbI$_3$, the materials were prepared via solution processing, the corresponding details can be found in the Supplementary Information. The performance and stability losses commonly observed in NiO$_x$-based PSCs could be due to either intrinsic reactivity of perovskite light absorber on NiO$_x$, or poor charge transport across the NiO$_x$/perovskite interface leading to the accumulation of electronic and ionic charge carriers, which would destabilize the light absorber. Therefore, it is necessary to experimentally deconvolute the origin of this issue, which has been done here by systematically investigating the intrinsic stability of MAPbI$_3$ on solution processed bare NiO$_x$. To look at the inherent instability due to interface reactivity, a set of MAPbI$_3$ samples with different layer thickness were deposited on NiO$_x$ by varying the concentrations of the perovskite precursors (1, 0.5, 0.375, 0.25, and 0.1M) in the solvent dimethylformamide (DMF); the films were prepared using a conventional anti-solvent step with chlorobenzene. By investigating this sample series, it should be possible to probe the perovskite bulk properties as well as the NiO$_x$/perovskite interface.

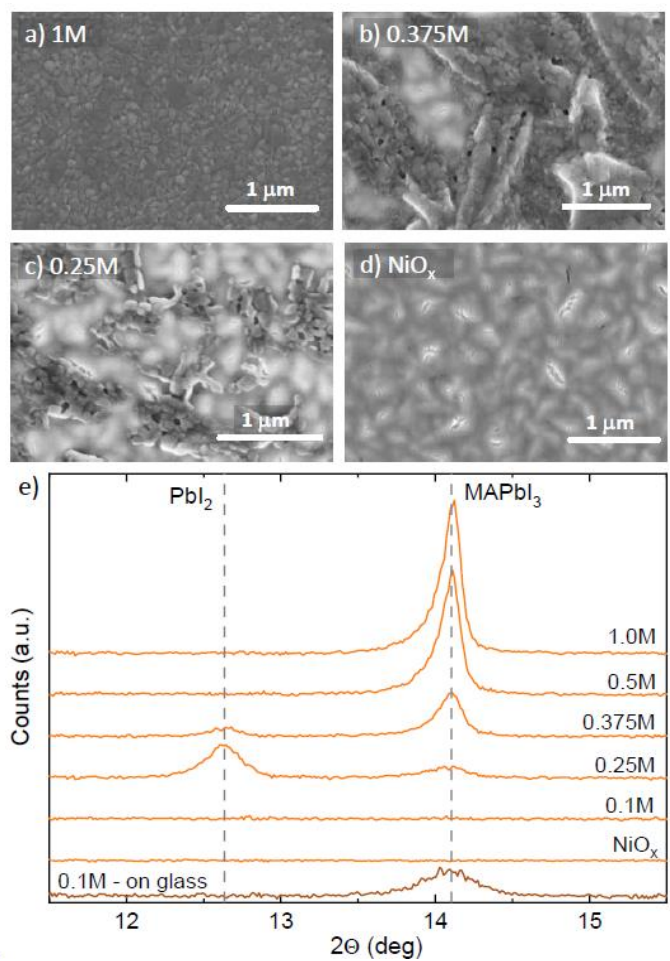

**Figure 1:** Investigation of the interface between NiO$_x$ and perovskite: SEM images of MAPbI$_3$ films from a) 1M b) 0.375M and c) 0.25M precursor solution concentrations deposited on d) a bare NiO$_x$ substrate; additional samples prepared from other MAPbI$_3$ concentrations are included in the Supplementary Information, Figure S1. d) XRD spectra of perovskite films prepared on NiO$_x$ using varying concentrations of precursor solution concentration.

The as-prepared solution processed bare NiO$_x$ as well as the perovskite films deposited on top were analyzed by SEM, the corresponding images are presented in Figure 1 a-d as well as Figure S1 in the Supplementary Information, where a larger data set is shown. The MAPbI$_3$ sample prepared from 1M precursor solution shows a compact, pin-hole free, uniform surface morphology with grain sizes of a few 100 nm, typical for this material. For concentrations of 0.5M and below, the bare NiO$_x$ surface becomes visible, meaning that the deposited films are non-uniform. Notably, for 0.5M and 0.375M the grain morphology still resembles the one observed for the thick film, however for lower concentrations, the appearance changes and the grains look less well-defined. This already indicates the decomposition of the thin film in contact to NiO$_x$. Moreover, a reduction of the precursor solution concentration to 0.1M (Figure S1 in the Supplementary Information) resulted in an SEM image looking similar to the bare NiO$_x$ surface, without traces of either grains or amorphous material.

The effect on the crystal structure of the perovskite formed on NiO$_x$ was analyzed by XRD measurements. The XRD traces of the samples prepared from the varying MAPbI$_3$ precursor solution concentrations on bare NiO$_x$ are summarized in Figure 1e. These spectra were obtained with minimal air exposure (< 10 min) and show the 2θ region relevant for the MAPbI$_3$ (110) lattice spacing at 14.1° as well as the PbI$_2$ (001) plane at 12.6°. Perovskite samples prepared from 1M and 0.5M precursor solutions displayed only the characteristic MAPbI$_3$ reflection at 14.1°, indicating the chemical integrity of the perovskite on NiO$_x$ for these thick layers. Similar measurements for 0.375M and 0.25M precursor solution-based samples revealed a less intense MAPbI$_3$ signal at 14.1° along with the emergence of the degradation product PbI$_2$ at 12.6°. The significant increase in the PbI$_2$ intensity from the 0.375M to the 0.25 M based sample evidences the degradation of the perovskite in contact to NiO$_x$. For the thinnest film with 0.1M concentration, no reflections are found, indicating either poor wettability of the perovskite solution on this surface or a complete degradation of the perovskite into amorphous material. To exclude the short air exposure during XRD or issues in film formation for the dilute solutions – especially in combination with an antisolvent treatment – as reasons for the formation of PbI$_2$, a control sample was prepared by depositing 0.1M perovskite precursor solution on a glass substrate. The corresponding XRD measurement is included in Figure 1e and notably shows the characteristic perovskite signal without any trace of PbI$_2$. From this difference it is clear that the reactivity issues observed in contact to NiO$_x$ are specific to this substrate and are likely emerging from the interface interactions between NiO$_x$ and MAPbI$_3$.

Further insights in the stability of MAPbI$_3$ on NiO$_x$, and possible interfacial interactions, are obtained from surface sensitive XPS measurements by analyzing the elemental composition and their respective oxidation states for the MAPbI$_3$ films deposited at different layer thickness on the NiO$_x$ substrates. The N1s core-level signals are shown in Figure 2 while the corresponding Pb4f and I3d signals are included in the Supplementary Information, Figure S2. For the MAPbI$_3$ layer prepared from 0.1M solution, signals for N, I and Pb can be observed. This is in contrast to the SEM and XRD measurements, where no traces of the material could be identified. Accordingly, only a very thin amorphous layer sticks onto NiO$_x$. By comparing the relative areas and binding energy positions of the observed core level peaks, we can estimate the composition of this layer, which is listed in Table 1. It seems that only approximately one quarter of the film is present as perovskite for the 0.1M deposition, while the rest is associated with adsorbed or reacted degradation products. More than 20% of the layer consists of an excess I$^-$ signal, which must be bound to the surface, as will be further discussed for the DFT calculations below. Intriguingly, this is in contrast to our previous study of organic halide precursors in contact with the metal oxide MoO$_3$, where low iodide concentrations were observed due to the formation of volatile I$_2$.[14] For the lead signal, shown in Figure S2, an additional lead peak at higher binding energy is observed close to the interface. This additional feature is difficult to assign, since rather inconsistent reports are found in literature with respect to binding energies. Due to the nature of the surface and the higher binding energy, we expect this to be PbO or $PbO_2$.[35,36] For this sample, the N peak shows a feature at lower binding energy, indicating that either the degradation product methylamine or another amine is physiosorbed at the interface.[13,37]

With increasing thickness, the composition of the film changes as can be seen from Table 1. For the 0.25M film, the additional nitrogen amine signal vanishes, instead a peak at higher binding energy appears, which is likely a nitrate.[13] Why it was not observed at lower perovskite precursor concentrations, closer to the interface, is unclear; this could be due to the rather low signal-to-noise ratio which makes fitting the data challenging. Also, for the lead measurement the previously observed additional Pb signal vanishes, indicating that both reaction products are confined to the very interface. Looking at the overall elemental composition, there are still excess I$^-$ surface bonds (18%) and additional $PbI_2$ (around 14%), so the perovskite is partially degraded, in agreement with the XRD measurements in Figure 1e. From 0.375M on, the peak intensities saturate and all side products become less pronounced, indicating that the surface measured here is mostly compromised of intact perovskite. For the 0.5M film, 90% of the surface consist of perovskite.

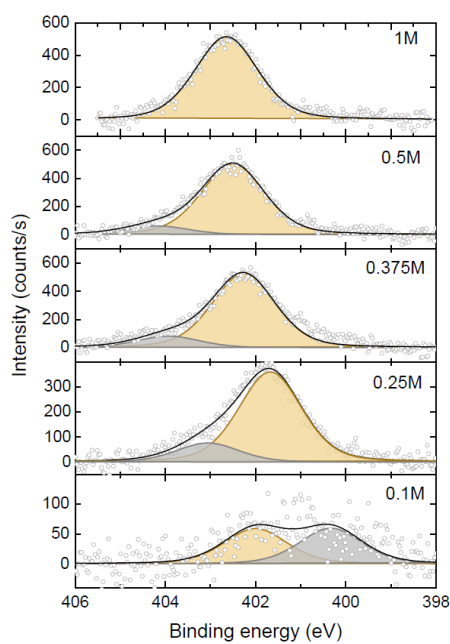

**Figure 2:** Investigation of the interface between $NiO_x$ and perovskite: XPS measurements of the N1s region of perovskite deposited on bare $NiO_x$; different molar concentrations were used as indicated in the Figure. The data was fitted by Voigt profiles, here the yellow shaded peaks correspond to the perovskite, while the grey peaks indicate the presence of nitrogen in a different chemical environment. The corresponding plots for Pb and I are shown in the Supplementary Information Figure S2.

**Table 1**: XPS-derived composition of the differently concentrated $MAPbI_3$ films deposited on bare $NiO_x$. Based on the measured N, Pb, and I peak intensities and their respective binding energies, different surface species have been identified.

| concentration (M) | $MAPbI_3$ (%) | $PbI_2$ (%) | Additional I surface bond (%) | Additional N species (%) | Additional Pb species (%) |
|---|---|---|---|---|---|
| 0.1 | 24 | 11 | 27 | 24 | 15 |
| 0.25 | 54 | 14 | 18 | 14 | 0 |
| 0.375 | 84 | 0 | 0 | 16 | 0 |
| 0.5 | 87 | 0 | 0 | 13 | 0 |
| 1 | 92 | 10 | 0 | 0 | 0 |

Overall, it can be concluded that the SEM, XRD, and XPS measurements carried out on MAPbI$_3$ samples prepared on bare NiO$_x$ consistently show that NiO$_x$ significantly affects the film formation, in particular for the more diluted concentrations, and that the perovskite clearly degrades in contact to NiO$_x$. While XRD can only detect the crystalline phase of the PbI$_2$ degradation product, XPS is also sensitive to the iodide surface bonds and additional degradation/reaction products which are likely amorphous. The strong chemical interactions between perovskite and NiO$_x$ are clearly demonstrated and can be expected to affect solar cell operation and stability.

### 2.1.2. NiO$_x$ surface passivation

In literature, the NiO$_x$ surface is reported to have a rich surface chemistry with significant concentrations of Ni(OH)$_2$, NiO(OH), Ni$_2$O$_3$ and non-negligible amounts of Ni$^{2+}$ vacancies as well as higher charged Ni$^{\delta+}$ states.[10,34,38] It is possible that a specific surface species is responsible for the reactivity of the perovskite film with this metal oxide, instead of the NiO itself. For instance, in the past we have shown that it is the presence of oxygen vacancies that triggers redox reactions in the case of MoO$_3$.[14] It is therefore of interest to identify the degradative reaction centers for NiO$_x$ and explore ways to passivate them in order to stabilize the perovskite. Many compounds such as metal salts, organic amines, and other electron donating materials have been reported to be effective in passivating the NiO$_x$ surface[19] and improving $V_{oc}$, PCE, and operational stability of the corresponding PSCs. To explore this, we selected a set of target-specific passivating materials that are shown in Scheme 1; similar metal cations,[39] halide anions, amines,[30,40] and hydrogen donors [34] have been reported to passivate NiO$_x$ surfaces. In this work, the selected materials are classified into three categories: (i) cation donors – metallic salts with mono or divalent cations which may potentially occupy Ni$^{2+}$ vacant sites in the NiO$_x$ surface, and corresponding anions to interact with high valent Ni centers; (ii) neutral bases (Lewis bases) – a neutral organic molecule with a capacity to donate electrons to the Ni atoms with higher formal charge (>2+); and (iii) hydrogen donors and/or anionic bases – a set of organic molecules that could donate protons to undercoordinated O atoms next to Ni$^{2+}$ vacancies and anions to interact with Ni$^{>2+}$. Identifying the nature of the interaction in a systematic study should make it possible to understand the degradative perovskite reaction pathways.

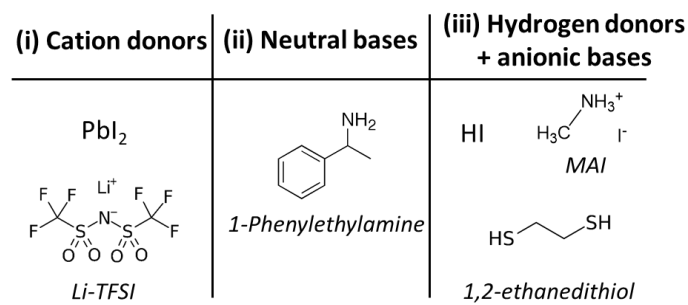

**Scheme 1**: Chemical structures of the categorized surface passivation materials for NiO$_x$.

In contrast to previous reports, here we are using a consistent passivation procedure for a large number of compounds, as shown in Scheme 1. In this process, the passivating organic materials were dissolved in either isopropanol or DMF (only in the case of PbI$_2$) to prepare a 0.1M solution; this was spin-cast on the NiO$_x$, followed by an annealing step at 100 °C for 15 min under inert condition. The heating process helps to overcome the reaction energy barrier and facilitates reaction kinetics between the passivating materials and NiO$_x$. This was followed by thorough rinsing of the substrates in the same solvent in which the respective passivating material was dissolved, ensuring a complete removal of unreacted or weakly bound material from the surface. This step ensures that the existing passivating materials at the interface are firmly bound to the NiO$_x$ surface. The presence of the passivating agents on the surface was confirmed by XPS, as shown in Figure S3 in the Supplementary Information. There, the I3d$_{3/2}$ signal is shown for HI, MAI, and PbI$_2$ as well as the S2p signal for 1,2-ethanedithiol / Li-TFSI, and the N1s for 1-phenylethylamine. Though the signals are weak in some cases, the measurements demonstrate the presence of bound passivating materials on NiO$_x$ even after thorough washing, which confirms the high chemical affinity of NiO$_x$.

We tried to identify changes in the NiO$_x$ surface composition due to chemical passivation through qualitative and quantitative assessment of XPS measured Ni2p$_{3/2}$ and O1s core-level peaks. These signals were collected from each of the treated NiO$_x$ substrates, and are shown in Figure S4 and S5 in the Supplementary Information. It is worth noting that fitting the Ni2p and O1s signals from NiO$_x$ is challenging. For the Ni2p peak, often a complicated de-convolution is made into signals corresponding to the different defect related oxidation states, e.g. NiO, Ni(OH)$_2$, Ni$_2$O$_3$, NiO(OH), as well as higher charged Ni$^{\delta+}$, as for example presented by Boyd et al.[34] However, in addition to the possible presence of these different surface species, the Ni2p peak for ionic compounds is known to have a complicated shake-up satellite structure.[41,42] Neglecting these pronounced satellite peaks leads to a significant overestimation of the presence of Ni components with higher oxidation states. As we are currently unable to separate the two effects, we choose not to present a fitting of the Ni2p peak in this work. Nonetheless, these measurements show that the different passivating agents had no significant effect on the shape of the Ni2p core level signal, as presented in Figure S5 in the Supplementary Information, suggesting that most of these Ni features at higher binding energy are indeed associated with shake-up structures. The O1s signal on the other hand is not affected by shake-up satellites, therefore the presence of different surface bonds can be determined from our data. As indicated in Figure S4 in the Supplementary Information, we identified 4 distinct sites, namely O in the NiO lattice, O in Ni(OH)$_2$, O bound to Ni$^{3+}$ and O from surface adsorbents, and under-coordinated O located next to the Ni$^{2+}$ vacant sites (vide infra).[43–46] Here, the effect of surface passivation can be deduced by comparing the quantitative change in the oxygen components with respect to the different passivating materials, the values are summarized in Table 2.

**Table 2**: Relative concentration of the different oxygen species, derived from the fitting of the O1s spectrum for bare as well as surface treated NiO$_x$ samples (data in Supplementary Information, Figure S4).

| Samples | O in NiO (%) | O in Ni(OH)$_2$ (%) | O @ Ni$^{3+}$ or surface adsorbed O (%) | O @ Ni$^{2+}$ vacancy sites (%) |
|---|---|---|---|---|
| NiO$_x$ | 54 | 33 | 5 | 8 |
| NiO$_x$ + PbI$_2$ | 60 | 35 | 2 | 2 |
| NiO$_x$ + Li-TFSI | 60 | 32 | 2 | 6 |
| NiO$_x$ + 1-phenylethylamine | 61 | 33 | 2 | 4 |
| NiO$_x$ + HI | 60 | 34 | 7 | 0 |
| NiO$_x$ + MAI | 57 | 36 | 7 | 0 |
| NiO$_x$ + 1,2-ethanedithiol | 63 | 34 | 4 | 0 |

Compared to the bare NiO$_x$ surface, a lower concentration of undercoordinated O atoms (i.e O @ Ni$^{2+}$ vacancy sites) are found for all treated samples. Notably, the treatment with the hydrogen containing compounds led to the complete disappearance of this O signal associated with Ni vacancies. No systematic variation in the relative Ni(OH)$_2$ concentration or the surface adsorbed oxygen species was observed. Therefore, based on the XPS data, the significant change in O concentration next to Ni$^{2+}$ vacancies suggests that treatments by MAI, HI and 1,2-ethanedithiol could produce the strongest change in the NiO$_x$ surface chemistry.

To test the resulting effect on the stability of the perovskite film, 0.25M MAPbI$_3$ precursor solutions were deposited onto the surface treated NiO$_x$ substrates. A selection of SEM images of these films are shown in Figure 3 a-d (for full sample set see Figure S6 in the Supplementary Information) and indicate an improved surface coverage in all cases compared to the same deposition on bare NiO$_x$. This observation clearly suggests that NiO$_x$ surface passivation improves the surface wettability of the perovskite precursor solution for all compounds investigated here, which could originate from the changes in chemical interactions between the passivated surface and MAPbI$_3$. Notably, for all treated substrates the MAPbI$_3$ morphology resembles a typical non-degraded perovskite layer.

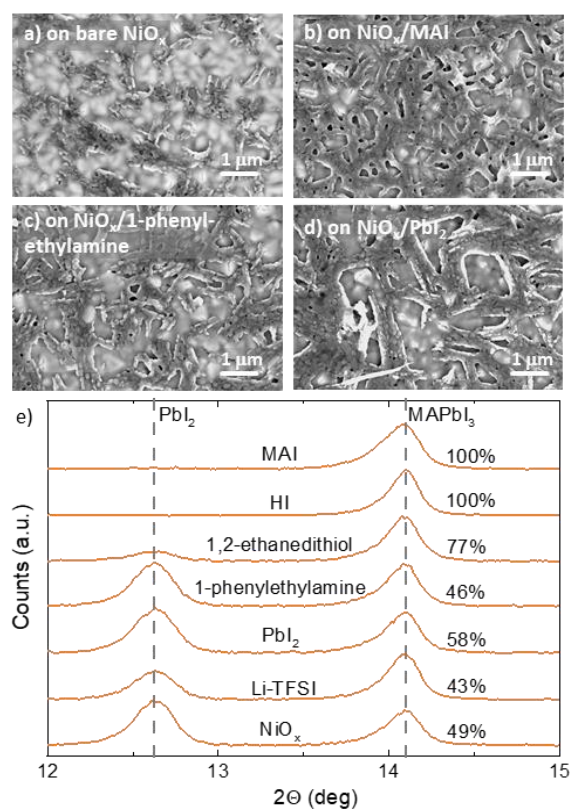

**Figure 3**: Effect of passivating agents on perovskite film formation. Sub-figures a) – d) show a selection of SEM images of MAPbI$_3$ films prepared from 0.25M precursor solution on bare and surface passivated NiO$_x$; additional images are included in the Supplementary Information Figure S6. e) XRD measurements of the same samples; the percentage of MAPbI$_3$ relative to PbI$_2$ signal is indicated in the Figure.

The XRD traces in Figure 3e also show remarkable changes in film stability. The values listed in the Figure indicate the percentage of intact MAPbI$_3$ signal relative to the PbI$_2$ degradation product. While the bare NiO$_x$ reference and most of the treated samples display less than 50% of MAPbI$_3$, this changes significantly for surfaces treated with the protic compounds. For MAPbI$_3$ on top of the 1,2-ethanedithiol treated NiO$_x$, this fraction increases to almost 80% and samples deposited on MAI and HI treated surfaces only show the peak corresponding to perovskite, indicating its complete chemical integrity on this surface passivated NiO$_x$. This observation suggests that protons are passivating defects/reaction centers which are otherwise leading to the degradation of MAPbI$_3$. Correlating these findings with the XPS derived oxygen species found for MAI, HI and 1,2-ethanedithiol treated NiO$_x$ indicates that the absence of undercoordinated O at the Ni$^{2+}$ vacant sites as the reason for MAPbI$_3$ stabilization, suggesting these O atoms as the perovskite degradative reaction center.

Further, given the proton donating ability, i.e. Bronsted acidity (p$K$a), of 1,2-ethanedithiol (8.96),[47] MAI (10.6),[18] and HI (-9.3),[48] all these substrates should have shown fully stable MAPbI$_3$, however, thiol seems to be slightly less effective. This suggests that in addition to proton abstraction a further reaction is responsible for fully passivating the surface, which could be interactions of the NiO$_x$ surface with the anionic halides (Lewis bases). A similar observation has been reported for solution prepared SnO$_x$ substrate, as Sn$^{4+}$ lattice sites on the surface interact preferably with halides rather than with thiol functional groups, which then passivate the surface.[49] In this study, this is further supported by two observations: i) excess iodide has been observed by XPS on the bare NiO$_x$ upon interacting with MAPbI$_3$ and ii) except for the proton containing compounds, the only significant increase in MAPbI$_3$ concentration is seen for the PbI$_2$ treated substrate. Thus, a complete NiO$_x$ surface passivation occurs through an initial proton abstraction by the reactive O atoms followed by iodide counterion reaction with neighboring Ni atoms. A similar concerted reaction was also proposed by Boyd et al. for MAPbI$_3$ degradation on NiO$_x$ before. [34]

From this data we can conclude that all these passivating materials do occupy the NiO$_x$ surface by chemically binding to reactive surface sites. Exceptional chemical and morphological stability of MAPbI$_3$ on NiO$_x$ is observed here for HI and MAI, while a decent stabilization was achieved for 1,2-ethanedithiol and a minor improvement for PbI$_2$. This suggests that the hydrogen donating ability of the chemicals is the decisive property when it comes to suppressing the perovskite degradation. While this follows a Bronsted acid-base reaction mechanism, a halide ion is also necessary for effective NiO$_x$ surface passivation. We can also identify defects on the NiO$_x$ to be responsible for initiating the MAPbI$_3$ degradation, since we correlated the undercoordinated oxygen atoms sitting next to the Ni$^{2+}$ vacancies as the effected binding site. The formal charges of these oxygens deviate from the bulk crystal value, possibly leading to the MAPbI$_3$ degradation on NiO$_x$[44] as will be discussed next.

### 2.1.3. DFT studies on the NiO surface

To explain the decomposition of MAPbI$_3$ on the NiO$_x$ surface at an atomistic scale, we have employed DFT first principal calculations; computational details can be found in the Supplementary Computational Details. From the XPS results reported above, an increased concentration of I$^-$ has been observed on the bare NiO$_x$ surface, which indicates that MAPbI$_3$ decomposes and degradation products stay adsorbed on the NiO$_x$ surface. Based on previous reports,[14,16] the MAI part of the perovskite is considered to be responsible for degradation and was therefore investigated here, instead of the complete perovskite structure. The decomposition of MAI on NiO surface can proceed through two pathways: deprotonation and dissociation, as represented in Figure 4.

For the investigation, we adsorbed MAI on pristine NiO, as well as surfaces with either an oxygen vacancy (NiO$_{vac\_O}$) or a Ni vacancy (NiO$_{vac\_Ni}$) to mimic the reported rich surface chemistry of the non-stoichiometric NiO$_x$ surface. Note that we have not modeled the Ni(OH)$_2$ surface, since in Table 2 this value did not vary significantly and therefore did not scale with the overall interface reactivity. In addition, the Ni vacant surface was passivated using either hydrogen (NiO$_{vac\ Ni}^{H}$), HI (NiO$_{vac\ Ni}^{HI}$), or PbI$_2$ (NiO$_{vac\ Ni}^{PbI2}$); see Figure S7 in the Supplementary Information for the atomistic representation. Table 3 lists the reaction energies and respective adsorption energies on these surfaces for the intact MAI (i.e. CH$_3$NH$_3$I) as well as the degradation products; the optimized coordinates are mentioned in the Supplementary Computational Details.

It is evident that MAI binds strongly to all surfaces and the same is the case for the degradation product HI; in both cases I$^-$ is adsorbed on a surface Ni atom, as seen in Figure 4. The high adsorption energy explains the observation by the XPS measurements presented above where a significant excess of iodide has been detected at the interface (see Table 2). Other degradation products also show negative adsorption energies, though the values are significantly lower, which could lead to the formation of volatile compounds in vacuum or during an XPS measurement.

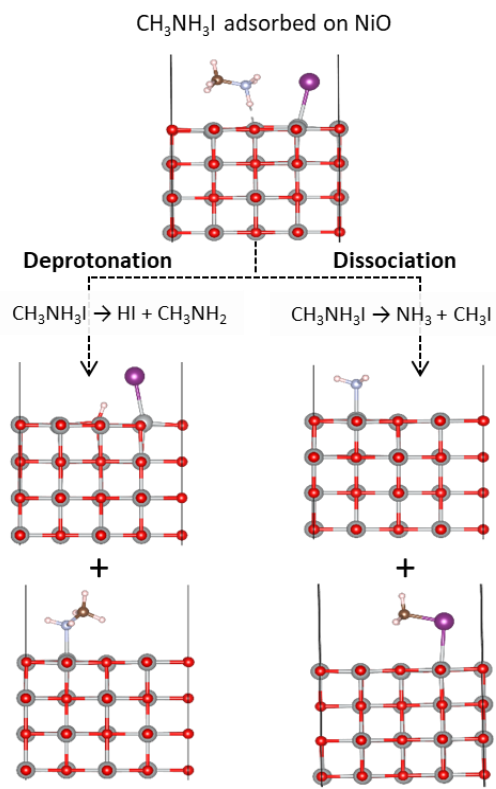

**Figure 4.** Schematic diagram of the MAI degradation reactions on a NiO surface: deprotonation (left) and dissociation (right). Red color balls indicate oxygen atoms, silver Ni atoms, violet iodine, brown carbon, white hydrogen, and light blue nitrogen.

Looking at the reaction energies, it is apparent that deprotonation is strongly favored over dissociation; the latter one is at a rather high positive value and will therefore not be further considered in the following discussion. For the deprotonation reaction we find a slightly negative value (-0.18 eV) for the pristine NiO surface, which becomes more negative in the presence of oxygen vacancies (-0.28 eV) and even more so for nickel vacancies (-0.49 eV). The fact that surfaces with $Ni^{2+}$ vacancies have the highest reactivity for deprotonation agrees well with our previous conclusion from the experimental study that the passivation of the undercoordinated oxygen located next to these $Ni^{2+}$ vacancies helps to stabilize the perovskite films. The DFT results for the H, HI, and $PbI_2$ passivated surfaces further help to understand the increased $MAPbI_3$ stability found in the XRD measurements presented in Figure 3. When the $Ni^{2+}$ vacancy is filled with a hydrogen atom, the reaction energy for deprotonation is significantly reduced to -0.18 eV, which is the same value as for the pristine NiO. If instead of a proton, HI is used for passivation, this value becomes highly positive (1.55 eV), meaning that the deprotonation reaction will not take place. This again is in excellent agreement with the experimental results, where we found that a proton transfer alone (as in the case of 1,2-ethanedithiol) is less effective in suppressing the degradation at the interface compared to the combined presence of a proton and an iodide (i.e. HI or MAI). For $PbI_2$ passivation, the deprotonation reaction is still possible (-0.10 eV), but to a lesser extent compared to NiO or defective NiO surfaces.

To figure out the reason why the HI treated surface performs so well in suppressing the deprotonation reaction, we have investigated the Bader charges on the Ni and O atoms in pure, Ni vacant, and HI treated surfaces; the detailed values can be found in the Supplementary Information, Figure S8. The calculations show an equal charge distribution of 1.2e and -1.2e on Ni and O atoms, respectively, in pristine NiO. When a $Ni^{2+}$ vacancy is present on the surface, the total charge present within Ni and O Bader volume is redistributed to maintain the charge neutrality of the system. While all Ni atoms display a slightly increased but similar Bader charge value of 1.24e, there are two types of oxygen atoms. An undercoordinated one near the $Ni^{2+}$ vacant site with lower Bader charge (-1.04e) and a fully coordinated one located away from the $Ni^{2+}$ vacant site with a slightly larger charge value (-1.16e). Such a difference in charge distribution on O atoms supports the peak fitting process performed on O1s signal from XPS measurements (see Figure S4 in Supplementary Information).

When introducing an HI molecule on this surface to investigate the reactive sites, the energy minimum is observed for the chemical configuration in which the proton is adsorbed on the undercoordinated O atom, despite a low electron charge on the latter, and the iodide ion on a neighboring Ni atom. Notably, as predicted above, such a deprotonation reaction initiated by the undercoordinated O atoms is the likely reason for $MAPbI_3$ degradation on $NiO_x$ surface. This suggests that the reactive undercoordinated O atoms as the origin of perovskite degradation interactions on NiO. This fact is further supported by the adsorption energy on pristine NiO surface for HI (-1.35 eV), which is lower compared to the case of a Ni vacancy (-1.98 eV). When this reactive O site is passivated by HI, the charge on the oxygen atom increases to -1.2e, which is equal to oxygen in the lattice, thus preventing further deprotonation by the same atom.

**Table 3**: Adsorption energies of MAI and its degradation products as well as reaction energies for deprotonation and dissociation on top of NiO, defective NiO, and passivated (Ni defective) NiO surfaces. $NiO_{vac\_O}$ and $NiO_{vac\_Ni}$ refers to NiO with nickel and oxygen vacancy, $NiO_{vac\_Ni}^{X}$ indicates the addition of a surface passivation "X" on a Ni defective NiO.

| Structure | Adsorption energy (eV) | | | | | |
| --- | --- | --- | --- | --- | --- | --- |
| | NiO | defective surfaces | | Ni defective surfaces with passivation | | |
| | | $NiO_{vac\_O}$ | $NiO_{vac\_Ni}$ | $NiO_{vac\_Ni}^{H}$ | $NiO_{vac\_Ni}^{HI}$ | $NiO_{vac\_Ni}^{PbI2}$ |
| MAI | -1.41 | -1.99 | -1.82 | -1.73 | -1.72 | -1.65 |
| $CH_3NH_2$ | -0.97 | -1.05 | -1.07 | -0.97 | 0.79 | -1.20 |
| HI | -1.35 | -1.96 | -1.98 | -1.68 | -1.69 | -1.30 |
| $NH_3$ | -0.78 | -0.89 | -0.94 | -0.85 | -1.11 | -1.02 |
| $CH_3I$ | -0.42 | -0.54 | -0.57 | -0.57 | -0.48 | -0.60 |
| Reaction energy | | | | | | |
| Deprotonation | -0.18 | -0.28 | -0.49 | -0.18 | 1.55 | -0.10 |
| Dissociation | 0.53 | 0.88 | 0.64 | 0.63 | 0.46 | 0.36 |

## 2.2. Perovskite device characterization

### 2.2.1. Surface treated $NiO_x$ as HTL in PSCs and their operational assessment

After investigating the stability of $MAPbI_3$ films on surface treated $NiO_x$ and unravelling the atomistic origin of the degradation, it is of interest to check the impact of these treatments on the performance of solar cell devices as well as their operational stability. A set of inverted p-i-n type PSCs were fabricated incorporating surface treated $NiO_x$ as HTL along with a reference device containing the bare $NiO_x$. For $NiO_x$ surface treatment, $PbI_2$, Li-TFSI, 1-phenylethylamine, HI, and MAI were selected as representative materials from each category of passivating compounds listed in Scheme 1. Apart from the change in $NiO_x$ surface treatment, all other constituting layers such as $MAPbI_3$ active layer, $C_{60}$ electron transport layer (ETL), bathocuproine (BCP) hole blocking layer, and Ag anode were kept the same for the entire series of devices. For each type of surface treatment, 14 devices were prepared, and their photovoltaic characteristics were extracted from the current density-voltage (*J-V*) measurements. The statistics on short-circuit current density ($J_{sc}$), open-circuit voltage ($V_{oc}$), fill factor (FF), and photon-to-current conversion efficiency (PCE) values of various surface treated $NiO_x$ containing working devices are compiled in Figure 5. The *J-V*-curves of champion devices and a more detailed listing of the device parameters are included in the Supplementary Information, Figure S9.

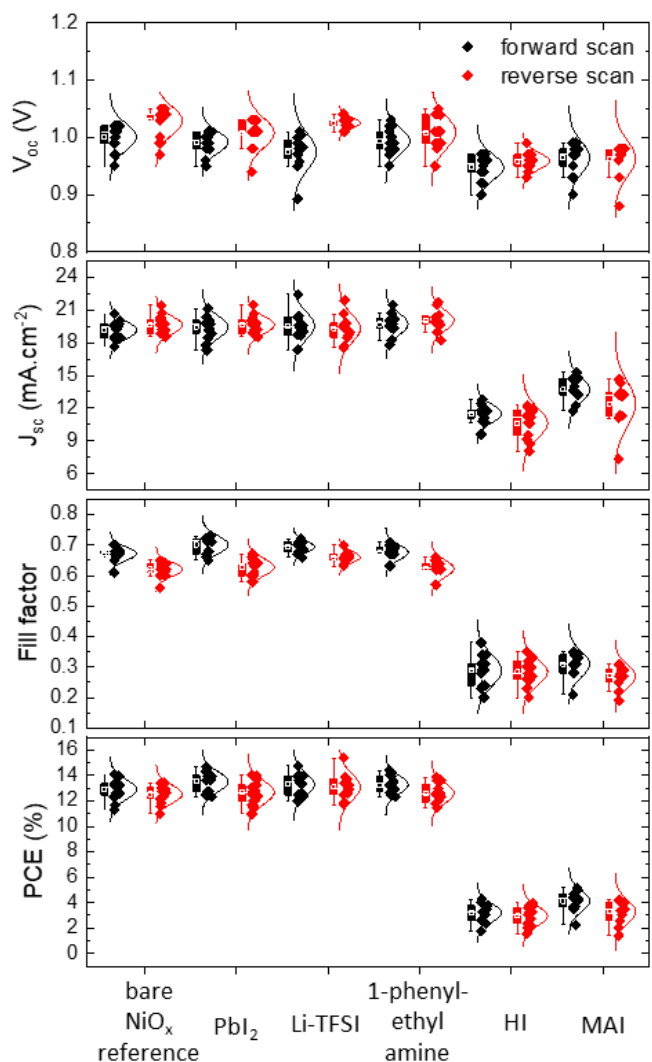

**Figure 5:** Photovoltaic characteristics of perovskite solar cells prepared on the bare $NiO_x$ as well as the surface treated $NiO_x$ hole transport layer. The values of $V_{OC}$, $J_{SC}$, FF, and PCE are derived from both reverse (red dots) and forward (black dots) scans at ca. 100 mV/s.

The champion reference device fabricated with bare $NiO_x$ as HTL displayed a maximum PCE of 14% with $V_{oc}$ of 1.04 V, $J_{sc}$ of 20.6 mA cm$^{-2}$, and FF of 0.67 for measurements of the *J-V* curves in the forward direction (from -0.2 V to 1.2 V). The same device in the reverse scan (1.2 V to -0.2 V) results in 1.01 V, 21.4 mA.cm$^{-2}$, 0.6, and 13.4%, respectively, suggesting only a small hysteresis in the device. On average, bare $NiO_x$ showed a PCE of approximately 13% and 12.6% in forward and reverse scans respectively, with only a small standard deviation of 0.8%. The solar cells with $PbI_2$, Li-TFSI, and 1-phenylethylamine treated $NiO_x$ as HTL show on average a similar performance to the bare reference devices (13.2% and 12.7% in forward and reverse scans, respectively). This observation is in line with the findings by XRD in Figure 3, which showed a similar $MAPbI_3$ content in these three cases. Therefore, there is no significant influence of these specific surface passivating materials in improving the PCE, though some of the champion devices surpass the best bare $NiO_x$ devices, see Table in Figure S9 in the Supplementary Information. Surprisingly, MAI and HI treated NiO$_x$ based devices, which showed superior stability for the MAPbI$_3$ thin films in Figure 3, display poor *J-V* characteristics with a significant S-shape behavior (Figure S9 in the Supplementary Information) and hysteresis. Further, the *J-V* curves were also evolving during subsequent scans without reaching a steady-state under the measured conditions. It should be noted that no voltage or light-based preconditioning was carried out on the investigated PSCs. The observed poor photovoltaic characteristics of these devices are indicative of significant ionic contributions to the measured current,[50,51] as a structural degradation of MAPbI$_3$ can be is ruled out. Consequently, the PCE values for HI and MAI passivated solar cells are in the range of approximately 4%, though reliable values cannot be given as the devices failed to reach a steady-state. Nevertheless, these results raise the question why HI and MAI treated NiO$_x$ surfaces lead to highly stable MAPbI$_3$ films, while at the same time displaying extremely poor solar cell performance.

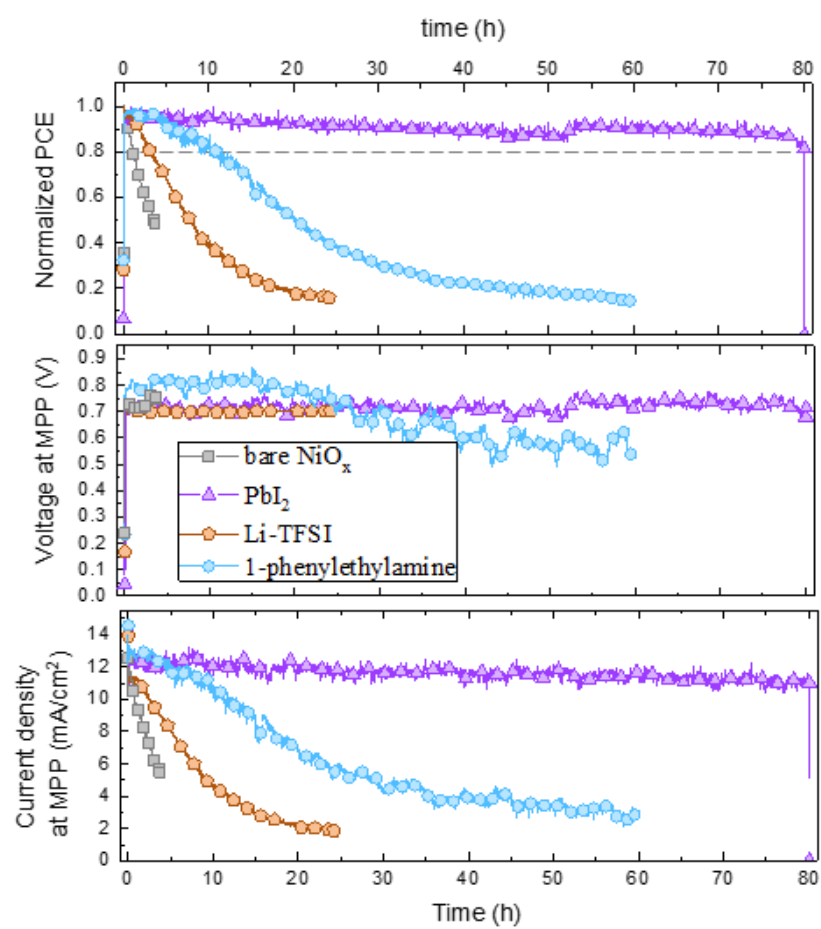

**Figure 6**: Operational stability results for the solar cells studied, as indicated. (a) Normalized power output with dashed line marking the 80% power limit to extract T$_{80}$ values, (b) voltage, and (c) current during MPP tracking under 0.8 sun equivalent illumination. For each surface treatment, the best performing solar cell with the longest T$_{80}$ lifetime is shown.

Further, the influence of NiO$_x$ surface treatment on solar cells operational stability was investigated by tracking the maximum power point (MPP) using the *Litos* system (Fluxim) under continuous white light illumination at 0.8 suns equivalent. For establishing the reliability of the measurements, at least 3 solar cells for each category of devices were measured. Figure 6 shows the power, voltage, and current output over time for the best-performing PSCs for each type of NiO$_x$ surface modification. Due to the low efficiency of the cells with MAI and HI, no stability tests were performed with Litos.

For the remaining solar cells, significant differences are found in device operational lifetime depending on the surface passivation. The device operational lifetime is characterized by the T$_{80}$ value, which is the time taken by a solar cell to reach 80% of their initial power output. Here, the reference NiO$_x$ and the Li-TFSI treated devices exhibit the shortest T$_{80}$ value of only one and three hours, respectively. The best operational stability is found for PSCs with 1-phenylethylamine and PbI$_2$ treated NiO$_x$, which displayed the T$_{80}$ value of 11 h and 80 h respectively. In all these devices, the performance drop is caused by a loss in current density as shown in Figure 6, while the MPP voltage remains rather stable. This clearly indicates the degradation of the MAPbI$_3$ film over time. See Table 3 for lifetime spans of the extended datasets.

**Table 3**: List of the range of T$_{80}$ lifetimes for at least 3 solar cells employing the different surface treatments. Note that the solar cells with MAI and HI treated NiO$_x$ could not be measured due to their low efficiency. Relative losses in solar cell parameters between fresh and T$_{80}$ device IV-scans are included in Table S1 in the Supplementary Information.

| HTL | range of T$_{80}$ [h] |
|---|---|
| bare NiO$_x$ | 1 |
| NiO$_x$ + PbI$_2$ | 30 - 80 |
| NiO$_x$ + Li-TFSI | 1 - 3 |
| NiO$_x$ + 1-phenylethylamine | 5 - 11 |

### *2.2.2. Characterization of ionic properties and recombination mechanisms*

It is clear that NiO$_x$ surface treatment influences the MAPbI$_3$ stability. However, it is puzzling that the improved thin film stability, introduced e.g. by the hydrogen donors, did not translate into more efficient or more stable solar cells. To gain a better understanding, we investigated the charge carrier dynamics within the devices in the microsecond time scale using the impedance spectroscopy (IS) and open-circuit voltage decay (OCVD) measurements. Here again, we compared bare NiO$_x$ HTLs with NiO$_x$ treated with one of the cation donors (PbI$_2$) and one of the neutral bases (1-phenylethylamine), while for the hydrogen donors we chose both HI and MAI since we wanted to get a better grasp on the extreme effect these surface treatments had on the solar cell performance.

Notably, in PSCs a method for the calculation of ion concentration has been introduced by Fisher et al.[52,53] combining the measurement of OCVD and photocurrent versus photo-voltage ($J_{sc} - V_{oc}$) curves under different illumination intensities.[54] To follow this approach, the $J_{sc} - V_{oc}$ curves were measured under varied illumination intensities and after long-term stabilization of the direct current (DC) mode signals for all devices, as shown in Figure S10 in the Supplementary Information. The OCVD data is presented in Figure 7a and was obtained by first stabilizing the initial value $V_{oc}(t = 0)$ under 0.2 sun equivalent white LED illumination. Then, the time dependent signal $V_{oc}(t)$ was measured in the dark at open-circuit condition after switching off the light source. This results in a fast initial voltage decay related to the electronic recombination of charge carriers and a subsequent slower reordering of mobile ions. The characteristic signal shape and decay time(s) are typically understood in terms of the photo-generated charge carrier density and recombination lifetime.[55] From the OCVD signal, a capacitance can be defined and integrated for estimating an effective mobile ion concentration [52] as:

$$N_{ion} \approx \frac{1}{qL} \int_0^{V_{oc}(0)} \bar{J}_{sc}(V_{oc}(t)) \left(\frac{\partial V_{oc}(t)}{\partial t}\right)^{-1} dV_{oc} \qquad (1)$$

where $q$ is the elementary charge, $L$ the perovskite layer thickness, $t$ is the time and $\bar{J}_{sc}$ is approached to the photocurrent corresponding to each photovoltage over time.

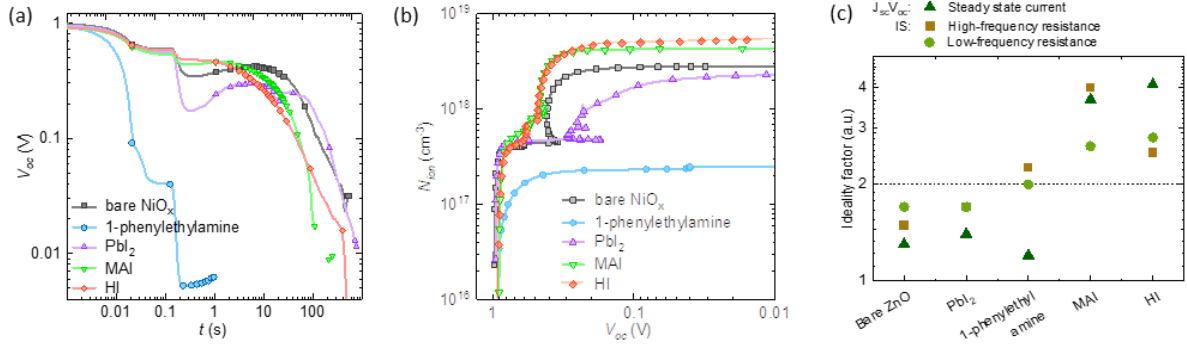

**Figure 7**: Open-circuit voltage decays (a) and corresponding integrated ionic concentration (b) for the studied samples, as indicated. (c) Combined values of ideality factors *m* determined by the different methods: photocurrent versus photovoltage under higher illumination intensities, as well as high and low frequency resistance from impedance spectroscopy from dark to high illumination intensities.

The calculated ion concentrations, based on Equation 1, were evaluated for each sample and are presented in Figure 7b, yielding values in the range of 2 - 5·$10^{18}$ cm$^{-3}$. Notably, the results of this method are obstructed by the practical limitations of extracting accurate values of the $J_{sc}$-$V_{oc}$ curve for low illumination intensities, over which most of Equation 1 is integrated. Nevertheless, clearly higher ion concentrations are found for the MAI and HI samples, which agrees with the pronounced hysteresis in the *J-V* curves for these devices in Figure S9 in the Supplementary Information. Notably, while for the stabilized $J_{sc}$-$V_{oc}$ curves an exponential behavior is found, this is not the case for the scan rate used to sweep the *J-V* curves in Figure S9. There could be various reasons for the substantial ion concentration determined in MAI and HI treated NiO$_x$ containing devices. The XPS measurements of the passivated surfaces in Figure S3 of the Supplementary Information show that compared to PbI$_2$, the treatments by HI and MAI lead to much higher iodide signals upon passivation, by a factor of 10 and 20, respectively. While the DFT calculations showed that HI addition to the surface is beneficial to fully suppress the deprotonation reaction of the perovskite, it is possible that some of the iodide can become mobile, leading to the measured high ion concentrations.

The significant increase of ion concentration for the MAI and HI samples suggest that ions build up towards the transport layers.[56] This is also supported by photoluminescence (PL) measurements displayed in Figure S11 in the Supplementary Information, which show a lower PL intensity from MAPbI$_3$ deposited on HI and MAI treated NiO$_x$, indicating predominant non-radiative charge recombination in the absorber layer influenced by the NiO$_x$/perovskite interface. In contrast, the devices with 1-phenylethylamine passivation agent showed a fast OCVD response, which suggests a lower mobile ion concentration within the framework of Fischer's model. This observation should be interpreted with care as Equation (1) neglects ion and electron mobility, which is closely linked to the ion concentration effects.[57] Thus, it can be hypothesized that 1-phenylethylamine passivation of NiO$_x$ could either decrease the mobile ion concentration in PSCs or increase the ion and/or electron mobility towards the HTL, or influences both processes in parallel.[58] Either way, the faster OCVD suggests that radiative recombination is favored in 1-phenylethylamine based samples , which is also supported by its intense steady state PL signal (Figure S11).

The analysis of the ideality factor *m* offers additional insights into differences among samples with varying dominant charge carrier recombination mechanisms. Typically,[53] the closer the value of *m* to unity, the higher the contribution of band-to-band radiative recombination over that of the Shockley-Read-Hall (SRH) trap-assisted nonradiative recombination. In this study, we utilize two different approaches for obtaining *m*. First, the steady state $J_{sc} - V_{oc}$ curves (Figure S10) measured under different illumination intensities were fitted to the empirical Shockley equation resulting in $J_{sc} \propto \exp[qV_{oc}/mk_BT]$, where $k_BT$ is the thermal energy. These steady-state values of *m* are included as green triangles in Figure 7c. However, the time-independency of this experiment hinders the separation between electron and ion properties. Alternatively, the ion-to-electron relation with respect to recombination and the ideality factors can be explored by IS at quasi-open-circuit condition under different illumination intensities.[59] IS utilizes the alternating current (AC) mode for obtaining the resistive and capacitive properties of the samples by measuring the current signal due to a small voltage perturbation. As a result, the characteristic response times, defect densities, and the fundamental built-in field can be assessed, among others.[60] Particularly, the main recombination mechanisms can be accessed by parameterizing the IS under different illumination intensities. Illustratively, the IS spectra for the studied samples are shown in Figure S12, along with the equivalent circuit model used for simulation and parameter extraction. Two main resistive elements, at high and low frequencies, were identified and parameterized to the equivalent circuit. The fitted high-frequency (Hf) and low-frequency (Lf) resistances as a function of $V_{oc}$ are presented in Figure S13, along with the exponential fittings to the behavior $R \propto \exp[-qV_{oc}/mk_BT]$.[54] In Figure 7c, these values of *m* determined from the Hf and Lf regions of the IS spectra are compared to the results from the steady state $J_{sc} - V_{oc}$ curves.

Values of 1 < *m* < 2 are found for the bare $NiO_x$ as well as the 1-phenylethylamine and $PbI_2$ treated $NiO_x$-containing solar cells. Specifically, the devices with bare $NiO_x$ and $PbI_2$ treated $NiO_x$ as HTL show a better agreement between the different techniques/conditions for the determination of the ideality factor. This aligns with the initial performance trends and the similar integrated values of $N_{ion}$ in Figure 7b: the higher the ionic effects the higher the value of *m* and the differences between techniques/conditions. In contrast, the 1-phenylethylamine treated $NiO_x$ containing cell shows a large discrepancy between the steady-state result from the $J_{sc} - V_{oc}$ and those from the time-resolved IS analysis. This could also be due to significant differences in the ion/electron mobility with respect to the other samples,[56–58,61] as discussed above. Moreover, the MAI and HI samples presented *m* > 2 under all experimental conditions with ideality factors as high as *m* ≈ 4. This not only suggests major SRH recombination but also a strong interface recombination effect, possibly related to ion accumulation at the interface and field screening.[57] Notably, high values $m > 2$ are typically associated with resistive issues at an interface and/or leakage defects that hinder the rectifying behavior in many types of semiconductor junctions.[62–64] In fact, this is in line with the measured steady-state *J-V* curves from these devices as shown in Figure S9 of the Supplementary Information. However, the ionic migration features of PSCs complicate the electrical response by including extra ion accumulation towards the interfaces, which modify the internal electric field in which the electronic charges move. [56,59,65,66]

The following reasoning can be considered to understand the above argument. The effective splitting of the quasi-Fermi levels at open-circuit $\Delta E_{Fnp} = qV_{oc}^{in}$ is not equal to the experimentally measured value of the photovoltage $V_{oc}$. The relation between them is $V_{oc} = V_{oc}^{in}/\gamma$, with $\gamma > 1$ being a dimensionless parameter related to interface recombination and ion accumulation effects observed in PSCs. For larger values of $\gamma$, the experimentally estimated value of the ideality factor $m = \gamma m^{in}$ will also be higher. Here, $1 < m^{in} < 2$ is the "internal" ideality factor corresponding to the section of the device which is "a diode", apart from the interfaces and/or ion accumulation layers in series connection. Now, given this direct relation between *m* and $\gamma$, the calculated high *m* values (>2) for the MAI and HI treated $NiO_x$ based devices indicate that the value of $\gamma$ should be higher than those from the other samples for reproducing the anomalously high apparent ideality factors. In other words, the internal photovoltage in the MAI and HI samples is likely diminished by the interface recombination.

Notably, even though the general trend of the ideality factors from the IS was not fundamentally different from that of the steady-state *$J_{sc}$-$V_{oc}$* data, the differences between Hf and Lf values in the AC method are particularly interesting. For instance, in the devices with bare $NiO_x$ and 1-phenylethylamine or $PbI_2$ treated $NiO_x$ as HTL, the electronic Hf ideality factor from IS approached the corresponding ionic-influenced Lf value, and exceeded the values from $J_{sc}$-$V_{oc}$ data. This may suggest device instability between one measurement and the other and/or an ionic reordering during the measurement of the DC signal of the $J_{sc}$-$V_{oc}$, which is different to that of the IS. In other words, the steady-state reordering of mobile ions favors radiative recombination in these samples, whereas bias perturbation hinders the relaxation and hence enhances non-radiative recombination. In contrast, the MAI and HI samples not only show systematically high values of $m > 2$, but also a variety of behaviors when comparing DC and AC techniques and frequency ranges in the IS. The high $m$ values are a direct consequence of the low FF and the limited rectifying behavior of these devices, and thus limited assessment of ideality factors in general. Nevertheless, for instance, comparing Lf ideality factors from IS ($m \approx 2.8$) with those from $J_{sc}$-$V_{oc}$ data ($m \approx 3.8$) one can infer that the higher values from the steady-state technique are artifacts due to the different ion distributions at short-circuit and open-circuit, which is not necessarily related to those in operation condition. For this reason, the hysteresis of the MAI and HI treated $NiO_x$ containing PSCs is significantly higher when compared to the other cells as shown in Figure S9 of the Supplementary Information.

## 3. Conclusions

In summary, $MAPbI_3$ deposited on bare $NiO_x$ is found to be degrading spontaneously at the interface as shown by XRD, SEM, and XPS measurements, which corroborates the high surface reactivity of this metal oxide. To understand and possibly overcome this issue, $NiO_x$ surface passivation was carried out with a variety of materials. Protic compounds such as MAI, HI and 1,2-ethanedithiol are found to be effective in structurally stabilizing $MAPbI_3$ on $NiO_x$, whereas other passivating materials display only slight improvements compared to bare $NiO_x$. The detailed analysis of the surface passivated $NiO_x$ substrates by XPS in combination with DFT calculations identified undercoordinated oxygen located at the vicinity of $Ni^{2+}$ vacant sites as the origin of $MAPbI_3$ degradation. Proton passivation of such reactive surface oxygen suggests a Bronsted acid-base type reaction mechanism for $NiO_x$ surface passivation. It is essential that this reaction should be followed by a Lewis base-acid reaction between Ni with >2+ oxidation state and halide counter anions for complete structural stability of $MAPbI_3$ thin films on $NiO_x$, by which the deprotonation reaction can be fully suppressed.

PSCs fabricated with moderately stable $MAPbI_3$ on passivated $NiO_x$ - achieved by using $PbI_2$, Li-TFSI, and 1-phenylethylamine treatment - were showing similar PCE values as reference devices built on bare $NiO_x$. Nonetheless, the $T_{80}$ value – i.e. the operational stability - increased from approximately 1h for bare NiOx to ~80h for $PbI_2$ treatment, showing the promising effect of surface passivation. Surprisingly, solar cells with structurally fully stable $MAPbI_3$, achieved by HI and MAI passivation, displayed extremely poor photovoltaic characteristics with severe hysteresis. The observed discrepancy between structural $MAPbI_3$ stability and PSCs operational stability is due to an increased mobile ion concentration in the devices, as determined by IS and OCVD measurements. This is accompanied by high ideality factors, suggesting a significantly increased non-radiative charge carrier recombination possibly at the MAI and HI passivated $NiO_x$/perovskite interface, leading to poor FF and $J_{sc}$ values for the corresponding PSCs. The reason for the increased mobile ions concentration in these samples is likely due to the weakly bound iodide at the $NiO_x$ surface after MAI and HI treatment.

We have established that there can be a discrepancy between high structural stability of a perovskite film and high operational stability. Towards the goal of highly stable $NiO_x$-based solar cells we need to identify a surface treatment that is able to passivate the reactive undercoordinated O sites at the surface by donating protons without increasing the overall ion concentration. It can be expected that a better understanding of the complex nature of the $NiO_x$ surface properties and the development of targeted surface passivation strategies will help to design and fabricate highly performing and stable PSCs.

## Acknowledgements

J.M. and S.O. thank the Ministry of Economic Affairs Innovation, Digitalization and Energy of the State of North Rhine-Westphalia for funding under the grant SCALEUP (SOLAR-ERA.NET Cofund 2, id: 32). We acknowledge the Ministerio de Ciencia e Innovación of Spain, Agencia Estatal de Investigación (AEI) and EU (FEDER) under grants PID2022-140061OB-I00 and PCI2019-111839-2 (SCALEUP). O.A. thanks the Spanish State Research Agency (Agencia Estatal de Investigación) for the Juan de la Cierva 2021 Fellowship grant. S. J., A. G. and B. R. acknowledge the Swiss Federal Office of Energy (SFOE) for project grant SI/501958-01 (SCALEUP). M. C. T. acknowledges support by the Nancy and Stephen Grand Technion Energy Program (GTEP), the Israel Ministry of Energy under the grant SCALEUP (SOLAR-ERA.NET), the Israel National Institute of Energy Storage (INIES), and COST Action 18234 supported by COST (European Cooperation in Science and Technology).